\documentclass[amsthm]{ifacconf}
\usepackage[round,sort&compress]{natbib}        

\usepackage[T1]{fontenc}
\usepackage{microtype}
\usepackage{amsmath, amssymb, amsfonts}
\usepackage{mathtools}
\usepackage{tikz,pgfplots,subfig}

\usepackage[applemac]{inputenc}

 \pgfplotsset{compat=newest}

\newcommand{\ie}{\emph{i.e.},}
\newcommand{\eg}{\emph{e.g.},}
\newcommand{\mb}{\mathbf}
\newcommand{\mc}{\mathcal}
\newcommand{\mbb}{\mathbb}

\renewcommand{\vec}{\mathbf}

\DeclarePairedDelimiter{\interiorpars}{(}{)}
\newcommand{\interior}{\operatorname{interior}\interiorpars}

\begin{document}
%

\begin{frontmatter}
\title{Actuation attacks on constrained linear systems: a set-theoretic analysis}

\author[First]{P. A. Trodden}
\author[Second]{J. M. Maestre}
\author[Third]{H. Ishii}

\address[First]{Department of Automatic Control \& Systems Engineering, University of Sheffield, Sheffield, UK {\tt\small p.trodden@sheffield.ac.uk}}

\address[Second]{Departmento de Ingeniería de Sistemas y Automática, University of Seville, Seville, Spain {\tt\small pepemaestre@us.es}}

\address[Third]{Department of Computer Science, Tokyo Institute of Technology, Yokohama, Japan {\tt\small ishii@c.titech.ac.jp}}
\begin{abstract}%
This paper considers a constrained discrete-time linear system subject to actuation attacks. The attacks are modelled as false data injections to the system, such that the total input (control input plus injection) satisfies hard input constraints. We establish a sufficient condition under which it is not possible to maintain the states of the system within a compact state constraint set for all possible realizations of the actuation attack. The developed condition is a simple function of the spectral radius of the system, the relative sizes of the input and state constraint sets, and the proportion of the input constraint set allowed to the attacker.
\end{abstract}%


\end{frontmatter}
\section{Introduction}%
\label{sec:Introduction}%

The security of control systems to cyber-attacks has
become a pressing issue, owing to the ubiquity of computers and
networks and the vulnerabilities that these
introduce~\citep{Smith15}. In the context of feedback control,
attention has focused on several salient aspects of the cyber-security
problem, including attack detection, 
synthesis and the analysis of control system stability
and performance under different classes of attack, including denial of
service (DoS), deception and false data injection (FDI)~\citep{PDB13,TSS+15}.

In this paper, we study a simple instance of an actuation attack
problem---a type of data injection attack---and, using set-theoretic
methods, develop fundamental conditions under which it is not possible
to robustly defend the system. In particular, we consider the problem
of maintaining the states of a constrained linear system within a
given state target set while it is subject to
adversarial input disturbances. We consider that the input constraint
set is partitioned, via a scaling factor, into two portions: the
control input is selected from one portion, and the attack input
from the other, such that the overall input applied is
constraint admissible. The main result of the paper is the
characterization of a lower bound on the constraint scaling factor
such that robust stabilization of the system---and infinite-time
robust constraint satisfaction---for all realizations of the attack is
not possible.

We note that although the state-feedback setting is simpler than that
typically considered in the cyber-security literature---and renders
certain aspects of the problem, such as stealth and detection,
trivial---the results we obtain offer some insights into the relative
ease of attack a system according to its dynamics and constraints. The
developed bound on the scaling factor depends, in a natural way, on
the open-loop stability of the system, via its spectral radius, and
the relative shapes and sizes of the input and state constraint
sets. Following intuition, the bound confirms that more unstable
systems with smaller target sets are easier to attack, in that the
proportion of the input constraint set required by the attacker is
smaller, which may have implications for the signal power and energy
required for a successful attack.

A few other papers have used set-theoretic techniques in the context
of cyber-security. \citet{LSF16} propose a receding-horizon control law
utilizing robust reachability sets in order to mitigate FDI and DoS
attacks. \citet{MVM+10} and \citet{MS12} use reachability analysis in
order to characterize the impact of FDI attacks. The most closely
related work, however, appears to be from outside of this literature:
\citet{SSC17} considered a constrained linear (open-loop stable)
autonomous system subject to additive disturbances selected from
scaled disturbance set, and developed lower and upper bounds on the
critical scaling factor at which robust infinite-time constraint
satisfaction is not possible. The setting considered in the present
paper is, however, more general and the techniques employed are
necessarily different in order to handle the possibility of open-loop
instability of the system.

\emph{Notation:} The sets of non-negative and positive
reals are denoted, respectively, $\mbb{R}_{0+}$ and $\mbb{R}_+$. The
set of natural numbers, including zero, is $\mathbb{N}$. For
$a, b \in \mbb{R}^n$, $a \leq b$ applies element by element. For
$X, Y \subset \mbb{R}^n$, the Minkowski sum is
$X \oplus Y \triangleq \{ x + y: x \in X, y \in Y \}$; for
$Y \subset X$, the Minkowski difference is
$X \ominus Y \triangleq \{ x \in \mbb{R}^n: Y + x \subset X \}$. For
$X \subset \mbb{R}^n$ and $a \in \mbb{R}^n$, $X \oplus a$ means
$X \oplus \{a\}$. $AX$ denotes the image of a set
$X \subset \mbb{R}^n$ under the linear mapping
$A : \mbb{R}^n \to \mbb{R}^p$, and is given by
$\{ Ax : x \in X\}$. The set
$-X \triangleq \{-x: x \in X\}$ is the image of $X$ under reflection
in the origin. The support function of a non-empty set
$X \subset \mbb{R}^n$ is
$h_X(x) \triangleq \sup \{x^\top z : z \in X \}$. A C-set is a convex
and compact (closed and bounded) set containing the origin; a PC-set
is a C-set with the origin in its interior.


\section{Problem statement}%
We consider a discrete-time, linear time-invariant system,
\begin{equation}
x_{k+1} = Ax_k + Bu_k, \qquad k \in \mbb{N},
\label{eq:sys}
\end{equation}
where $x_k \in \mbb{R}^n$ and $u_k \in \mbb{R}^m$ are the state and input at time $k$. The states and inputs are
constrained as,
\begin{equation*}
x_k \in X \ \text{and} \ u_k \in U, \qquad k \in \mbb{N}.
\end{equation*}
\begin{assum}\label{assump:basic}
  The pair $(A,B)$ is reachable. $X$ and $U$ are PC-sets; $U$ is symmetrical about the origin, \ie~$U = - U$.
  \end{assum}

  The setting considered in this paper is that the
  system~\eqref{eq:sys} is subject to attacks on its input. We
  suppose that these attacks take place via an attacker gaining access
  to, and injecting data into, the control input signal, $u$. The
  system under attack is
\begin{equation}
  x_{k+1} = Ax_k + B(v_k + a_k)
  \label{eq:dyn}
  \end{equation}
  \ie~the input to the system is $u_k = v_k + a_k$, where
  $a_k \in U$ is the attack signal, and $v_k \in U$ the
  control signal provided by the system controller (the
  defender). More specifically, we consider that the attacker is able
  to use a proportion $\alpha \in [0,1)$ of the input constraint
  space, while the defender is left with the remaining proportion
  $1-\alpha$. For $ k \in \mbb{N}$,
  \begin{equation*}
    a_k \in \alpha U \ \text{and} \  v_k \in (1-\alpha) U.
\end{equation*}

In this way, the overall input constraint---which typically represents
a hard actuation limit---is respected, yet the attacker is able to
disturb the system and simultaneously reduce the set of actions
available to the defender.

The goal of the attacker is to drive the system states out of
$X$. The goal of the defender is, naturally, to keep the state
in $X$, despite the actions of the attacker. Our aim in this paper is to analyse this simple scenario and develop
fundamental conditions, in terms of the parameter $\alpha$, on when an
attack is \emph{undefendable}, \ie~such that there exists no control
law able to maintain the state within $X$.

\section{Preliminary analysis}

We begin with some definitions relevant to the problem, and then link
these to known concepts and results in constrained control. The
following refer to the system~\eqref{eq:dyn} and constraints
$(x_k,v_k,a_k) \in X \times (1-\alpha)U \times \alpha U$.

\begin{defn}[Attack and defence sets and strategies] \ \\
  The \emph{admissible attack \{defence\} set} is $\alpha U$
  \{$(1-\alpha)U$\}, with $\alpha \in [0,1)$. An \emph{admissible
    attack \{defence\}} is an action $a_k \in \alpha U$
  \{$v_k \in (1-\alpha)U$\}. An \emph{admissible attack \{defence\}
    strategy} is a policy $x \mapsto v \in \alpha U$
  \{$x \mapsto a \in (1-\alpha) U$\}.
  \end{defn}

\begin{defn}[Undefendable and defendable attack set] \ \\
  An attack set $\alpha U$ is said to be \emph{undefendable} for
  the system~\eqref{eq:dyn} if, for all $x_0 \in X$, there does
  not exist an admissible defence strategy that maintains
  $x_k \in X$ for all $k \geq 0$. Otherwise, an attack set is
  said to be defendable.
\end{defn}

There is a direct link and equivalence between these definitions and established concepts in
the literature: infinite reachability, strong reachability and
robust control invariance~\citep{Bertsekas72, Blanchini99, Kerrigan00}.

\begin{defn}[\citet{Bertsekas72}]
  A set $Y \subset R^n$ is said to be:
  \begin{enumerate}
  \item \emph{Infinitely reachable} if there exists a control
    law $\mu(\cdot)$ and some $x_0 \in Y$ such that $x_k \in Y$ and $u_k = \mu(x_k) \in (1-\alpha) U$ for all $w_k \in \alpha U$.
    \item \emph{Strongly reachable} or \emph{robust control invariant} (RCI) if there exists a control
    law $\mu(\cdot)$ such that for all $x_0 \in Y$, $x_k \in Y$ and $u_k = \mu(x_k) \in (1-\alpha) U$ for all $w_k \in \alpha U$.
       \end{enumerate}
  \end{defn}

  The link to defendability follows trivially.

  \begin{lem}
    The attack set $\alpha U$ is defendable if, and only if, $X$ is
    infinitely reachable. $X$ is infinitely reachable if, and only if, it contains a robust control invariant set $C$.
    \end{lem}

    \begin{rem}
      In establishing a link between these concepts and the results
      later in the paper, we make a tacit assumption on the
      information pattern in the problem: the defender selects $v_k$
      with knowledge of $x_k$ but without knowledge of $w_k$, while the
      attacker may have knowledge of both $x_k$ and $v_k$.
      \end{rem}
    
    This motivates the remainder of the paper. The question of whether
    an attack set is defendable or undefendable (as these concepts are
    defined) amounts exactly to whether or not the state constraint
    set $X$ contains an RCI set. If it does, then the attack set can
    be said to be defendable\footnote{Defendability, as it is
      defined, is a weak notion, in the sense that it does not imply
      that all $x_0 \in X$ can be kept within $X$.} and standard
    techniques from robust constrained control can be used to keep the
    state within $X$. If it does not, then an attack set is
    undefendable, and there does not exist any defence strategy that
    keeps the state within $X$ for all time, accounting for \emph{all}
    possible actions of the attacker\footnote{Undefendability
      says nothing about how the attacker may determine an admissible
      attack strategy that achieves the goal of steering $x$ outside
      $X$. This is a less standard control problem than that
      of the defender, and is beyond the scope of the paper.}. Our more concrete aim
    is, therefore, to characterize the relation between the constraint
    scaling factor $\alpha$ and the existence of an RCI set within
    $X$.

\section{Robust constraint-admissible and control invariant sets }

First we present some known results and new results regarding RCI sets, with respect to a general linear system
 \begin{equation}
   \begin{aligned}
  x_{k+1} &= Ax_k + Bu_k + Ew_k,\\
  (x_k,u_k,w_k) &\in X \times U \times W.
  \end{aligned}\label{eq:sysg}
  \end{equation}
  These are subsequently specialized to the setting described in the
  previous section.

  \subsection{Some known results}

  The $i$-step robust constraint-admissible set is the set of all states
  that can be kept within $X$ for at least $i$ time steps, for any disturbance, respecting
  the input constraints:
\begin{multline*}
C_i := \bigl\{ x : \exists \mb{u}_i \in \mc{U}_i \ \text{such that} \ \mb{x}_i \in \mc{X}_i  \ \text{for all} \ \mb{w}_i \in \mc{W}_i  \bigr\}
\end{multline*}
where $\mb{u}_i$ (respectively $\mb{w}_i$) is the sequence of $i$
controls $\{u_0,u_1,\dots,u_{i-1}\}$ (disturbances
$\{w_0,w_1,\dots,w_{i-1}\}$), the set
$\mc{U}_i \triangleq {U} \times \dots \times {U}$, with a
similar definition for $\mc{W}$ and ${W}$. The corresponding
sequence $\mb{x}_i = \{x_0,x_1,\dots,x_i\}$ is obtained by, starting
from $x_0$, applying the input sequence $\mb{u}_j$ and disturbance
sequence $\mb{w}_j$. The definition requires
$\mb{x}_i \in \mc{X}_i \triangleq X \times \dots \times X$.

We recall some basic facts about $C_i$ and its limit
$C_\infty$~\citep{Bertsekas72,Blanchini94,Kerrigan00}:
\begin{lem}
  Suppose $U$ is a PC-set and $W$ is a C-set. Then (i)
  $C_0 = X$; (ii) if $X$ is compact [convex], then
  each $C_i$ is closed [convex]; (ii)
  $C_{i+1} \subseteq C_i$; (iii)
  $C_i = \bigcap_{j=0}^i C_j$; (iv)
  $C_{\infty} := \lim_{i \to \infty} C_i =
  \bigcap_{i=0}^{\infty} C_i$; (v) if
  $0 \in \interior{C_\infty}$ then every $C_i$ is a PC-set; (vi) if $0 \in \interior{C_{\infty}}$, then ${C}_\infty$ is a robust control invariant set for the system~\eqref{eq:sysg} and constraint set $(X,U,W)$; (vii)
  $C_{\infty} $, if non-empty, is maximal in the sense that it
  contains all other robust control invariant sets for the system~\eqref{eq:sysg}.
\end{lem}

To compute $C_i$, the following recursion holds:
\begin{align*}
  C_{i+1} &= Q(C_{i}) \cap X,\\
  \text{with} \ C_0 &= X,
\end{align*}
and where $Q(\cdot)$ is the backwards reachability operation:
\begin{equation*} Q(Y) \triangleq \{ x : \exists u \in U \
  \text{such that} \ Ax + Bu \oplus EW \subseteq Y \}.
\end{equation*}
More specifically, for the linear time invariant
system~\eqref{eq:sysg},
\begin{equation*}
C_{i+1} = (A)^{-1} \bigl([C_{i} \ominus EW] \oplus (-BU) \bigr) \cap X,
\end{equation*}
where $(A)^{-1}(\cdot)$ denotes the pre-image of the linear
transformation $A(\cdot)$, and exists regardless of whether $A$ is
invertible; for shorthand we will write $A^{-i} Y$ to denote $(A^i)^{-1}(Y)$.

\begin{lem}~
  \begin{enumerate}
  \item $C_\infty$ is finitely determined if and only if there exists an $i^* < \infty$ such that $C_{i^*+1} = C_{i^*}$.
  \item If $C_\infty$ is a PC-set, then such an $i^*$ exists.
\end{enumerate}
\end{lem}

\subsection{Some new results}

The aim of the paper is to characterize the existence of the set
$C_\infty$ in terms of $\alpha$. This requires the analysing of the
sequence of sets $\{C_i\}$, the dynamics of which are characterized by
Minkowski additions, subtractions, intersections and preimages, and not
readily amenable to analysis. The following result therefore, which
appears to be new, gives insight into how $C_i$ (and therefore
$C_\infty$) may be characterized in terms of sets with simpler
dynamics that are more amenable to analysis.
\begin{prop}\label{prop:Tdef}
  The set $C_{i}$ is bounded as
  \begin{align*}
    C_i &\subseteq \bigcap_{j=0}^{i} A^{-j} T_j 
               \intertext{where}
               T_{i+1} &= (T_i \ominus A^iEW) \oplus A^i(-BU)\\ 
    \text{with} \ T_0 &= X.             
  \end{align*}
  
\end{prop}

\begin{rem}\label{rem:tight}
  
  A special case of this result was reported by~\citet{SSC17}, who
  considered an autonomous system $x_{k+1} = Ax_k + Ew_k$ subject to a
  disturbance from a scaled set $\alpha W$. In that setting,
  \ie~without a control input, what we refer to here as ${C}_i$ is the
  $i$-step $0$-reachability set, and $C_\infty$ is the maximal robust
  \emph{positively} invariant set. The authors determine conditions on
  the scaling constant $\alpha$ under which $C_\infty$
  exists. \citet{SSC17} develop the following
  relation~\eqref{eq:Schulze}, which we show now to be a
  corollary of Proposition~\ref{prop:Tdef}.
   \begin{cor} If ${U} = \{0\}$ then ${T}_i = {X} \ominus {R}_i$ and
      \begin{equation}\label{eq:Schulze}
      {C}_i = \bigcap_{j=0}^i (A^j)^{-1}( {X} \ominus {R}_i).
\end{equation}
\end{cor}

Note that in~\eqref{eq:Schulze} the relation for $C_i$ holds with
equality, and not just the inclusion depicted in
Proposition~\ref{prop:Tdef}. The reason for the weakening of the
equality to mere inclusion is the behaviour of the Minkowski sum under
the intersection: for sets $A$, $B$ and $C$,
$(A \cap B) \oplus C \subseteq (A\oplus C) \cap (B \oplus C)$ and not
$(A \cap B) \oplus C = (A\oplus C) \cap (B \oplus C)$; we used this
fact in the proof of Proposition~\ref{prop:Tdef}. The latter equality
does hold if the \emph{union} of convex sets $A$ and $B$ is convex,
which is generally not the case. This, together with the fact that we
consider a more general setting in this paper anyway (\citet{SSC17},
because no control input is available, necessarily restrict their
developments to stable $A$) means that the methodologies and results
of~\citet{SSC17} do not apply.
\end{rem}

We conclude the section by establishing sufficient conditions for
emptiness of the set $C_i$ for some $i > 0$ and subsequently
$C_\infty$. The results are central to the developments in the
next section, when we specialize to the input-attack setting.

\begin{prop} \ \\
        If, for some $i^\star > 0$, $T_{i^\star} = \emptyset$ then
        $C_{i} = \emptyset$ for all $i \geq i^\star$.
      \label{prop:Cexist}
      \end{prop}

\begin{prop} \ \\
 If, for some $i^\star > 0$, $S_{i^\star} = \emptyset$, where
    \begin{equation*}
     S_i \triangleq X \oplus \left[\bigoplus_{j=0}^{i-2} A^j B(-U)\right] \ominus \left[\bigoplus_{j=0}^{i-1} A^j EW \right],
      \end{equation*}
      then $T_{i} = \emptyset$ for all $i \geq i^\star$.
      \label{prop:Texist}
      \end{prop}

\section{For which values of $\alpha$ is an attack set undefendable?}

We now recast some of the results from the previous section in the
particular setting of the paper, and develop conditions under which
$C_\infty$ does not exist. 

Specializing the definitions of $C_i$ and $T_i$ to the system~\eqref{eq:dyn}
and constraints
$(x_k,v_k,a_k) \in X \times (1-\alpha) U \times \alpha U$, and
exploiting the symmetry of $U$, we obtain
\begin{align*}
  C^\alpha_{i+1} &= A^{-1}([C_{i}^\alpha \ominus \alpha BU] \oplus (1-\alpha)BU) \cap X\\
  \text{with} \ C^\alpha_0 & = {X},
\intertext{and}
 {T}^\alpha_{i+1} &= ({T}^\alpha_i \ominus \alpha A^iB {U} ) \oplus (1-\alpha) (A^iB{U})\\
    \text{with} \ {T}^\alpha_0 &= {X},             
  \end{align*}
  where the sets are super-indexed by $\alpha$ to denote their
  dependency on this scaling factor. The connection between the two is, following Proposition~\ref{prop:Tdef},
  \begin{equation*}
    C_{i}^\alpha \subseteq \bigcap_{j=0}^i A^{-j} T^\alpha_j.
    \end{equation*}
    In a similar way, the set $S_i$ in Proposition~\ref{prop:Texist}
    may be specialized to the setting and denoted $S_i^\alpha$.

    Our goal is to determine, for each $i^\star \in \mbb{N}$, the
    smallest $\alpha$ for which $C_{i^\star}^\alpha$ is empty:
    \begin{equation*}
      \alpha_{i^\star} \triangleq \inf \{ \alpha : C_{i^\star}^\alpha = \emptyset, \alpha \in [0,1] \}.
    \end{equation*}

    In our main result, Theorem~\ref{thm:main}, we establish an upper
    bound on $\alpha_{i^\star}$. We achieve this by characterizing,
    for each $i^\star \in \mbb{N}$, an $\bar{\alpha}_{i^\star} $ that
    renders $S^{\alpha}_{i^\star}$ empty for all
    $\alpha > \bar{\alpha}_{i^\star}$. By
    Propositions~\ref{prop:Cexist} and \ref{prop:Texist}, any
    $\alpha > \bar{\alpha}_{i^\star} \geq \alpha_{i^\star}$ then
    ensures that $C^{\alpha}_{i^\star}$ is empty. We find that this
    bound depends on the relative sizes of the constraint sets $X$ and
    $U$, as well as the relative stability or instability (via the
    spectral radius) of the open-loop system.

    The following assumption is key to the development and simplicity
    of the result:

    \begin{assum}
    The dominant eigenvalue of $A$ is real and positive.
    \label{assump:eig}
  \end{assum}

  Let $\rho_A$ denote the spectral radius of $A$, and
  $\mathcal{V} \triangleq \{v_A^1,\dots,v_A^r, -v_A^1,\dots,-v_A^r\}$
  be the set of linearly independent eigenvectors corresponding to the
  dominant eigenvalue, plus their additive inverses. Define
  \begin{equation*}
  H_{XU}(\bar{v}_A) \triangleq \min_{v \in \mc{V}} h_X(v) / h_{BU}(v),
\end{equation*}
which is the smallest among the ratio of support functions to $X$ and $BU$
evaluated in the directions $\pm {v}^i_A$, $i=1\dots r$; define
$\bar{v}_A$ as the corresponding eigenvector. The following assumption
ensures this is well defined.

 \begin{assum}\label{assump:sup}
    The mapped set $BU$ has non-zero support in at least one of the directions $\pm {v}^i_A, i = 1 \dots r$.
    \end{assum}

  \begin{thm}\label{thm:main}
    Suppose Assumptions~\ref{assump:basic},~\ref{assump:eig} and \ref{assump:sup} hold. If, for some $i^\star \in \mbb{N}$,
    \begin{equation*}
      \alpha > \bar{\alpha}_{i^\star}\triangleq \begin{dcases} \frac{1 + H_{XU}(\bar{v}_A)[1-\rho_A] - \rho_A^{i^\star-1}}{2 - \rho_A^{i^\star-1} - \rho_A^{i^\star}} & \rho_A \neq 1 \\
        \frac{H_{XU}(\bar{v}_A) + i^\star -1}{2i^\star - 1}  & \rho_A = 1
        \end{dcases}
      \end{equation*}
      and $\bar{\alpha}_{i^\star} < 1$ then $C^\alpha_{i} = \emptyset$
      for all $i \geq i^\star$.
    \end{thm}

Two corollaries of this theorem follow.

\begin{cor}
  If $\alpha > \bar{\alpha}_{i^\star}$ when
  $\bar{\alpha}_{i^\star} < 1$ for some $i^\star \in \mbb{N}$, then the
  attack set $\alpha U$ is undefendable; moreover, the state is
  guaranteed to remain in $X$, for all attack strategies, for at most
  $i^\star-1$ steps.
\end{cor}

    \begin{cor}
     If 
      \begin{equation*}
        \alpha > \bar{\alpha}_\infty \triangleq
        \begin{dcases}
          \frac{1 + H_{XU}(\bar{v}_A)[1- \rho_A]}{2} & \rho_A < 1 \\
          \frac{1}{1 + \rho_A} & \rho_A \geq 1
          \end{dcases}
        \end{equation*}
and $\bar{\alpha}_{\infty} < 1$ then $C^\alpha_\infty = \emptyset$.
      \end{cor}
  
      The bounds obtained here provide insight into the relative ease
      of attacking a system depending on its dynamics and
      constraints. More specifically, the critical scaling factor
      depends on the most unstable eigenvalue of the system and the
      relative sizes of the state and input constraint sets in the
      direction of the corresponding eigenvector. The result implies
      that unstable systems are easier to attack (for example, if
      $\rho_A > 1$ then $\bar{\alpha}_\infty < 1/2$, so the attack set
      does not need to be as large as the defence set to render the
      system undefendable) and also that (un)defendability depends on
      the relative sizes of the sets $BU$ (the mapped inputs) and $X$
      (for example, even if $\rho_A = 0$, the system can be rendered
      undefendable if $\bar{\alpha}_\infty < 1$, which requires
      $H_{XU}(\bar{v}_A) < 1 \implies h_{BU}(\bar{v}_A) >
      h_X(\bar{v}_A)$).

      \begin{rem}
        It should be noted that Assumption~\ref{assump:eig} places
        restrictions only on the dominant eigenvalues. The proof (in the Appendix) reveals why: under this assumption,
        the long-term critical evolution of the set
        $A^i EW = \alpha A^i BU$, by which the intermediate sets are
        restricted, is in the direction $\bar{v}_A$, which enables the
        simple result obtained. It is possible to extend the result to
        more general $A$ matrices, such as those with complex dominant
        poles. However, because the long-term growth of the set
        $\alpha A^i BU$ is then not in single direction, it is more
        involved to determine the number of steps after which the set
        $T_i$ becomes empty.
      \end{rem}

      \begin{rem}
        Although the synthesis of attack strategies is
        beyond the scope of this paper, it is interesting to note that
        a simple strategy for unstable systems presents itself
        in light of the derived bound. Since
        $\bar{\alpha}_{\infty} < 1/2$ for all $\rho_A > 1$, the
        attacker can choose $\alpha = 1/2$ and---information
        pattern permitting---employ the attack strategy $a(x,v) = -v(x)$
        to guarantee that the state leaves $X$ in finite time.
        \end{rem}

\section{Illustrative examples}

We illustrate the results of the paper via three example systems:
\begin{align*}
  \mb{S}_1 : A = \begin{bmatrix} 0.5 & 1 \\ 0 & 0.7 \end{bmatrix} &&
  \mb{S}_2 : A = \begin{bmatrix} 1 & 1 \\ 0 & 1 \end{bmatrix} &&
  \mb{S}_3 : A = \begin{bmatrix} 1.9 & 1.1 \\ 0.5 & 1.5 \end{bmatrix}
  \end{align*}
  where, in each case,
  $B = \begin{bmatrix} 0.5 & 1 \end{bmatrix}^\top$. The sets $X$ and
  $U$ are the unit hypercubes.

\begin{figure}
    \centering\footnotesize
%
%
\begin{tikzpicture}

\begin{axis}[%
width=0.8\linewidth,
height=0.4\linewidth,
scale only axis,
xmin=-1.1,
xmax=1.1,
ymin=-1.1,
ymax=1.1,
xlabel = {$x_1$},
ylabel = {$x_2$},
legend style={legend cell align=left,align=left,draw=white!15!black}
]

\addplot[area legend,dashed,thick,draw=black,fill=none,fill opacity=0]
table[row sep=crcr] {%
x	y\\
0.670242931263323	-0.162822266017778\\
0.844357076780759	-0.409111933395006\\
0.96969696969697	-1\\
0.562420764380738	-1\\
-0.0430717800464722	-0.449561536729835\\
-0.438899517180246	-0.0784853169280581\\
-0.670242931263323	0.162822266017777\\
-0.844357076780759	0.409111933395005\\
-0.969696969696969	0.999999999999999\\
-0.562420764380736	0.999999999999999\\
0.0430717800464996	0.44956153672981\\
0.438899517180231	0.0784853169280732\\
}--cycle;

\addplot[area legend,solid,thick,draw=black,fill=none,fill opacity=0]
table[row sep=crcr] {%
x	y\\
0.616869905375742	-0.180917799682273\\
0.790614614534048	-1\\
0.562420764380737	-1\\
-0.0430717800465317	-0.449561536729781\\
-0.258323557760555	-0.247769675052494\\
-0.469617553781498	-0.0273750566326625\\
-0.616869905375742	0.180917799682273\\
-0.790614614534048	1\\
-0.562420764380736	1\\
0.0430717800465152	0.449561536729796\\
0.258323557760557	0.247769675052492\\
0.469617553781499	0.0273750566326616\\
}--cycle;

\addlegendentry{$\bigcap_{j=0}^3 A^{-j} T_j$}
\addlegendentry{$C_3$}

\end{axis}
\end{tikzpicture}%
    \caption{Comparison of $C_3$ and its bounding set $\bigcap_{j=0}^{3} A^{-j} T_j$ for system $\mb{S}_3$ with $\alpha = 0.1$.}
    \label{fig:1}
  \end{figure}
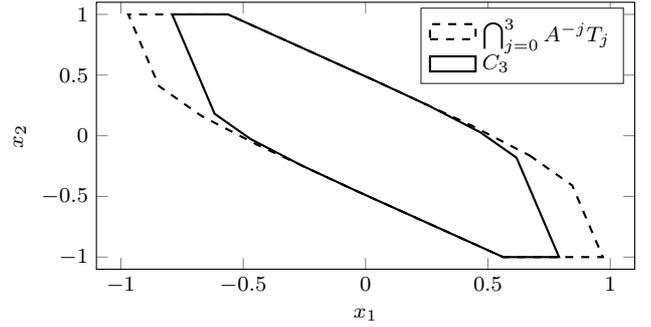

  First we illustrate Proposition~\ref{prop:Tdef}. Fig.~\ref{fig:1}
  compares, for system $\mb{S}_3$ and a scaling factor of
  $\alpha=0.25$, the three-step constraint admissible set $C_3$ with
  the outer bounding set $\bigcap_{j=0}^3 A^{-j} T_j$ derived in
  Proposition~\ref{prop:Tdef}. The inclusion is not tight, as pointed
  out in Remark~\ref{rem:tight}.

\begin{figure}
  \centering\footnotesize
  \subfloat[Stable system $\mb{S}_1$]{
%
%
\begin{tikzpicture}

\begin{axis}[%
width=0.8\linewidth,
height=0.3\linewidth,
scale only axis,
xmin=0,
xmax=19,
xlabel={$i^\star$},
ymin=0,
ymax=1.05,
ylabel={$\alpha_{i^\star}$ or $\bar{\alpha}_{i^\star}$},
]

\addplot [color=red,solid,mark=asterisk,thick,mark options={solid}]
  table[row sep=crcr]{%
2	0.845986984815618\\
3	0.765978207838673\\
4	0.721917680943975\\
5	0.694148181976587\\
6	0.67513338135429\\
7	0.661363053225825\\
8	0.650981120278521\\
9	0.642913855924956\\
10	0.636496883341387\\
11	0.631297007751806\\
12	0.627019660137845\\
13	0.623457517767121\\
14	0.620460404854859\\
15	0.617916854662079\\
16	0.615742381672669\\
17	0.613871777680001\\
18	0.612253908035458\\
19	0.610848109869534\\
};

\addplot [color=blue,thick,solid,mark=o,mark options={solid}]
  table[row sep=crcr]{%
2	0.63\\
3	0.604\\
4	0.59\\
5	0.581\\
6	0.575\\
7	0.57\\
8	0.567\\
9	0.564\\
10	0.562\\
11	0.561\\
12	0.559\\
13	0.558\\
14	0.557\\
15	0.556\\
16	0.556\\
17	0.555\\
18	0.554\\
19	0.554\\
};

\addlegendentry{$\bar{\alpha}_{i^\star}$}
\addlegendentry{${\alpha}_{i^\star}$}

\end{axis}

\end{tikzpicture}%
}\\
  \subfloat[Double integrator $\mb{S}_2$]{
%
%
\begin{tikzpicture}

\begin{axis}[%
width=0.8\linewidth,
height=0.3\linewidth,
scale only axis,
xmin=0,
xmax=19,
xlabel={$i^\star$},
ymin=0,
ymax=1.05,
ylabel={$\alpha_{i^\star}$ or $\bar{\alpha}_{i^\star}$},
]

\addplot [color=red,solid,mark=asterisk,thick,mark options={solid}]
  table[row sep=crcr]{%
2	0.8\\
3	0.714285714285714\\
4	0.666666666666667\\
5	0.636363636363636\\
6	0.615384615384615\\
7	0.6\\
8	0.588235294117647\\
9	0.578947368421053\\
10	0.571428571428571\\
11	0.565217391304348\\
12	0.56\\
13	0.555555555555556\\
14	0.551724137931034\\
15	0.548387096774194\\
16	0.545454545454545\\
17	0.542857142857143\\
18	0.540540540540541\\
19	0.538461538461538\\
};

\addplot [color=blue,thick,solid,mark=o,mark options={solid}]
  table[row sep=crcr]{%
2	0.462\\
3	0.441\\
4	0.44\\
5	0.436\\
6	0.426\\
7	0.418\\
8	0.412\\
9	0.408\\
10	0.406\\
11	0.405\\
12	0.404\\
13	0.403\\
14	0.402\\
15	0.402\\
16	0.401\\
17	0.401\\
18	0.401\\
19	0.401\\
};

\addlegendentry{$\bar{\alpha}_{i^\star}$}
\addlegendentry{${\alpha}_{i^\star}$}
\end{axis}

\end{tikzpicture}%
}\\
    \subfloat[Unstable system $\mb{S}_3$]{
%
%
\begin{tikzpicture}

\begin{axis}[%
width=0.8\linewidth,
height=0.3\linewidth,
scale only axis,
xmin=0,
xmax=19,
xlabel={$i^\star$},
ymin=0,
ymax=1.05,
ylabel={$\alpha_{i^\star}$ or $\bar{\alpha}_{i^\star}$},
]
\addplot [color=red,solid,mark=asterisk,thick,mark options={solid}]
  table[row sep=crcr]{%
1	0.557707162168758\\
2	0.380724831016468\\
3	0.323580117818578\\
4	0.302287912977706\\
5	0.293939333175376\\
6	0.290600810225683\\
7	0.289255275386115\\
8	0.2887112712945\\
9	0.288491048939157\\
10	0.288401853217747\\
11	0.288365719129306\\
12	0.288351079609597\\
13	0.288345148288984\\
14	0.288342745132858\\
15	0.288341771455637\\
16	0.288341376953808\\
17	0.288341217114559\\
18	0.288341152352894\\
19	0.28834112611357\\
};
\addplot [color=blue,thick,solid,mark=o,mark options={solid}]
  table[row sep=crcr]{%
1	0.492\\
2	0.326\\
3	0.295\\
4	0.289\\
5	0.286\\
6	0.272\\
7	0.261\\
8	0.253\\
9	0.247\\
10	0.243\\
11	0.239\\
12	0.236\\
13	0.234\\
14	0.232\\
15	0.231\\
16	0.229\\
17	0.228\\
18	0.227\\
19	0.226\\
};

\addlegendentry{$\bar{\alpha}_{i^\star}$}
\addlegendentry{${\alpha}_{i^\star}$}

\end{axis}

\end{tikzpicture}%
}
    \caption{Comparison of the bound $\bar{\alpha}_{i^\star}$ obtained from Theorem~\ref{thm:main} with the true bound $\alpha_{i^\star}$.}
    \label{fig:2}
  \end{figure}
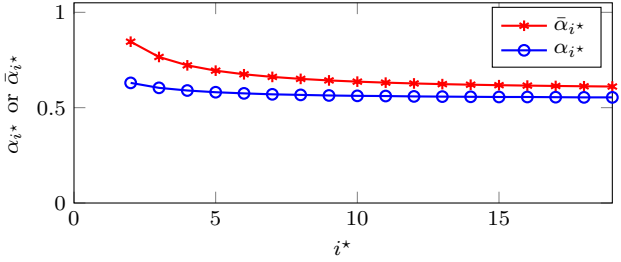
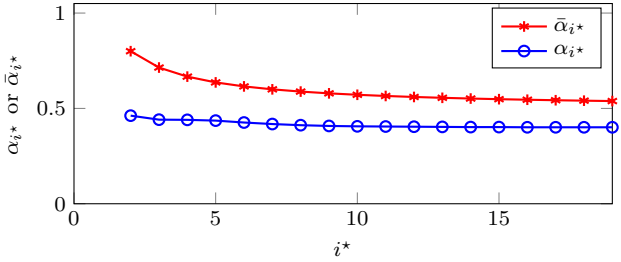
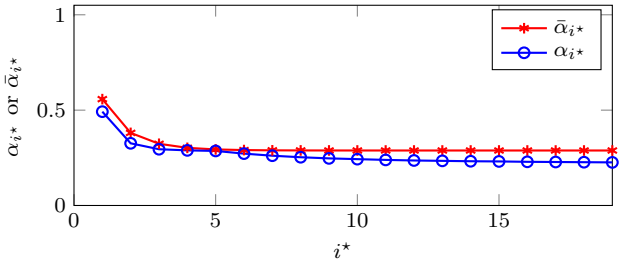

  Next, we illustrate and investigate the result of
  Theorem~\ref{thm:main} and its corollaries. Figure~\ref{fig:2}
  shows, for the systems $\mb{S}_1$ to $\mb{S}_3$, the exact critical
  scaling factors $\alpha_{i^\star}$ and the upper bound
  $\bar{\alpha}_{i^\star}$ established in Theorem~\ref{thm:main}. The
  exact scaling factors were found by trial and error, searching over
  $\alpha \in (0,1)$ until the smallest value is found for which the
  backwards reachability recursion
  $C^{\alpha}_{i+1} = Q(C_i^\alpha) \cap X$ results in
  $C^{\alpha}_{i} = \emptyset$ for $i = i^\star$.  

  \section{Discussion}

  The derived bound on critical $\alpha$ is merely sufficient and, as the numerical
  results indicate, it is not tight---$C_i^\alpha$ may
    become empty at some smaller $i$ or $\alpha$ than the bound of
    Theorem~\ref{thm:main} suggests. The sources of conservatism are
  threefold:
  \begin{enumerate}
  \item Proposition~\ref{prop:Cexist} is sufficient, but not
    necessary, to guarantee emptiness of $C_{i^\star}$ at some
    $i^\star$. The lack of necessity arises from the lack of tightness
    (illustrated in Fig.~\ref{fig:1}) in the inclusion relation
    between $C_i$ and $T_i$ established in
    Proposition~\ref{prop:Tdef}. This itself is, as explained in
    Remark~\ref{rem:tight}, because Minkowski addition is not
    distributive over set intersections. To avoid this, and guarantee equality in the
    relation of Proposition~\ref{prop:Tdef} and necessity of the
    condition in Proposition~\ref{prop:Cexist}, strong and unusual
    assumptions on the system and constraints would be required.
  \item Similarly, Proposition~\ref{prop:Texist} is merely sufficient
    to ensure emptiness of $T_{i}$ at a given $i = i^\star$. The
    source of conservatism again arises from the basic properties of
    Minkowski addition and subtraction, and in particular that these
    operations do not commute: Proposition~\ref{prop:Texist} uses the
    fact that, for sets $A$, $B$ and $C$,
    $A \oplus C \ominus B \supseteq A \ominus B \oplus C$.
  \item Finally, Theorem~\ref{thm:main} is sufficient, but not
    necessary, to ensure emptiness of $S_{i^\star}^\alpha$. The
    condition was developed by considering the dynamics of
    $S_i^\alpha$ over times $i \in \mbb{N}$ in \emph{only one}
    direction in $\mbb{R}^n$, the eigenvector associated with the
    dominant (real) eigenvalue. It is possible, but this depends on
    the system and constraints, that emptiness of $S_{i^\star}^\alpha$
    could be concluded at some smaller $\alpha$ than
    $\bar{\alpha}_{i^\star}$ by considering other directions.  On the
    other hand, the long-term change in $S_i^\alpha$ will tend to be
    dominated by activity in the direction of the dominant eigenvalue;
    our numerical results show relatively good agreement between the
    theoretical bound and the exact bound as $i^\star$ grows.
    \end{enumerate}
    
    Finally, it is worth remarking that the developed theoretical
    bounds are simple to determine, requiring only the spectral
    information about the open-loop system and a couple of support
    function evaluations on the constraint sets $X$ and $U$. In
    comparison, to determine the exact $\alpha(i^\star)$ requires a
    search over the space $\alpha \in (0,1)$, computing the sequence
    $\{C_i^\alpha\}_{i \geq 0}$ until it is found that $C^\alpha_{i^\star}$ is empty.

\section{Conclusions and future work}

We have analysed a simple instance of a constrained linear system
subject to actuation attacks and, using set-theoretic methods, derived
a lower bound on the sufficient size of the attack set in order that
robust infinite-time constraint satisfaction can not be
guaranteed. The bound depends, in an intuitive way, on the spectral
radius of the system and size and shape of the constraint sets. Future work will consider more general instances of $A$ (\eg~with dominant
complex eigenvalues) and a more sophisticated setting (\eg~with outputs and
sensor data injection).

\appendix

\section{Proof of Proposition~\ref{prop:Tdef}}

  We have $\mc{C}_0 = X = A^{-0}T_0$ and
  \begin{equation*}
    \begin{split}
      \mc{C}_1 &= A^{-1}\left( [X \ominus EW] \oplus (-BU) \right) \cap {X} \\
    &= A^{-1} \left( [T_0 \ominus EW] \oplus (-BU) \right)\cap A^{-0} T_0\\
    &= A^{-1} T_1 \cap A^{-0} T_0 .
    \end{split}
    \end{equation*}
    Now ${C}_2 = A^{-1} \left([{C}_1 \ominus EW] \oplus (-BU) \right) \cap X$. Denote $\mb{W} \triangleq EW$, $\mb{U} \triangleq (-BU)$, then
    \begin{equation*}
\begin{split}
  C_2 &= A^{-1} \left([ (A^{-1} T_1 \cap A^{-0} T_0) \ominus \mb{W}] \oplus \mb{U} \right) \cap X\\
  &= A^{-1} \left([ (A^{-1} T_1 \ominus \mb{W}) \cap (A^{-0} {T}_0 \ominus \mb{W})] \oplus \mb{U} \right) \cap {X}\\
  &\subseteq A^{-1} \left([ (A^{-1} {T}_1 \ominus \mb{W}) \! \oplus \! \mb{U}]  \cap [(A^{-0} {T}_0 \ominus \mb{W}) \! \oplus \! \mb{U}]\right)\! \cap \! {X} \\
  & = A^{-2}( {T}_1 \ominus A\mb{W} \oplus A\mb{U} ) \cap A^{-1}( {T}_0 \ominus \mb{W} \oplus \mb{U}) \cap {T}_0\\
  &= A^{-2} {T}_2 \cap A^{-1} {T}_1 \cap A^{-0} {T}_0,
  \end{split}
    \end{equation*}
    where the second line follows from
    $(X \cap Y) \ominus Z = (X \ominus Z) \cap (Y \ominus
    Z)$~\citep{KG98}, the third from $(X \cap Y) \oplus Z \subseteq (X \oplus Z) \cap (Y \oplus Z)$~\citep{Schneider93}, and the fourth from $A^{-1}(X \cap Y) = A^{-1} X \cap A^{-1} Y$. Thus, $C_2 \subseteq \bigcap_{j=0}^2 A^{-j} T_j$. Similar arguments establish the same for all $i > 2$. \qed

    \section{Proof of Proposition~\ref{prop:Cexist}}

    By Proposition~\ref{prop:Tdef}, $C_i \subseteq \bigcap_{j=0}^i A^{-j} T_j$.
            It follows that, if $T_{i^\star} = \emptyset$ for some
            $i^\star$, then $C_i = \emptyset$ for all
            $i \geq i^\star$. \qed

    \section{Proof of Proposition~\ref{prop:Texist}}

    Consider $T_{i+1} = (T_i \ominus A^i EW ) \oplus A^iB(-U)$. Applying this recursively from $T_0 = X$, we obtain
    \begin{multline*}
      T_{i} = X \ominus A^0EW \oplus A^0B(-U) \\ \ominus A^1EW \oplus A^1B(-U) \\ \vdots \\ \ominus A^{i-1} EW \oplus A^{i-1} B(-U)
      \end{multline*}
      For non-empty sets $A$, $B$ and $C$, we have $(A \ominus B) \oplus C \subseteq (A \oplus C) \ominus B$. Therefore, for all $i \geq 0$, 
      \begin{equation*}
        T_{i-1} \subseteq X \oplus \left[\bigoplus_{j=0}^{i-2} A^j B(-U)\right] \ominus \left[\bigoplus_{j=0}^{i-2} A^j EW \right]
        \end{equation*}
        Consider how $T_i$ is determined from $T_{i-1}$: first, $A^{i-1}EW$ is subtracted. It follows that
        \begin{equation*}
          T_{i^\star} = \emptyset \iff T_{i^\star-1} \ominus A^{{i^\star}-1}EW = \emptyset \ \text{for some} \ i^\star \geq 0.
        \end{equation*}
          The latter is true if (but not only if)
          \begin{equation*}
            X \oplus \left[\bigoplus_{j=0}^{i^\star-2} A^j B(-U)\right] \ominus \left[\bigoplus_{j=0}^{i^\star-2} A^j EW \right] \ominus A^{i^\star-1}EW = \emptyset,
            \end{equation*}
            rearrangement of which gives the condition in the hypothesis. \qed

\section{Proof of Theorem~\ref{thm:main}}

Consider
  \begin{equation*}
S_i^\alpha =    X \oplus (1-\alpha) \left[\bigoplus_{j=0}^{i-2} A^j BU\right] \ominus \alpha \left[\bigoplus_{j=0}^{i-1} A^j BU \right]
      \end{equation*}
      and take the support function of this set in the direction of
      the dominant eigenvector $\bar{v}_A$:
\begin{equation*}
  \begin{split}
h_{S_i^\alpha}(\bar{v}_A) &\leq h_X(\bar{v}_A) \\ &\quad + (1-\alpha) \sum_{j=0}^{i-2} h_{A^j BU}(\bar{v}_A) - \alpha \sum_{j=0}^{i-1} h_{A^j BU}(\bar{v}_A) .
\end{split}
\end{equation*}
By definition, for a set $Y$ and matrix $M$, $h_{MY}(z) = h_Y(M^\top z)$. Therefore, and re-arranging the summations,
\begin{multline*}
h_{S_i^\alpha}(\bar{v}_A) \leq h_X(\bar{v}_A) + (1-2\alpha) \sum_{j=0}^{i-2} h_{BU}((A^j)^\top \bar{v}_A) \\ - \alpha h_{BU}((A^{i-1})^\top \bar{v}_A) .
\end{multline*}
Since $\bar{v}_A$ is an eigenvector of $A$ corresponding to a real,
positive eigenvalue $\bar{\lambda}_A$,
$A^k \bar{v}_A = \bar{\lambda}_A^k \bar{v}_A$ for all $k$ in
$\mbb{N}$. Moreover, since $\bar{\lambda}_A$ is the dominant
eigenvalue, $\bar{\lambda}_A^k \bar{v}_A = \rho_A^k
\bar{v}_A$. Therefore,
$h_{BU}((A^{k})^\top \bar{v}_A) = \rho_A^k h_{BU}(\bar{v}_A)$ for all
$k \in \mbb{N}$. The previous inequality simplifies to
\begin{equation*}
  h_{S_i^\alpha}(\bar{v}_A) \leq h_X(\bar{v}_A) + \left[ (1-2\alpha)  \sum_{j=0}^{i-2} \rho_A^j - \alpha \rho_A^{i-1}\right] h_{BU}(\bar{v}_A).
\end{equation*}
Now consider the geometric series
$\sum_{j=0}^{i-2} \rho_A^j = 1 + \rho_A + \rho^2_A + \dots + \rho_A^{i-2}$. There are
two cases:
\begin{equation*}
  \sum_{j=0}^{i-2} \rho_A^j = \begin{cases} \frac{1 - \rho_A^{i-1}}{1 - \rho_A} & \rho_A \neq 1 \\
    i-1 & \rho_A = 1
    \end{cases}
  \end{equation*}
  Consider $\rho_A \neq 1$ first. We have 
  \begin{equation*}
  h_{S_i^\alpha}(\bar{v}_A) \leq h_X(\bar{v}_A) + \left[ (1-2\alpha) \frac{1 - \rho_A^{i-1}}{1 - \rho_A} - \alpha \rho_A^{i-1}\right] h_{BU}(\bar{v}_A).
\end{equation*}
It follows that $h_{S_{i^\star}^\alpha}(\bar{v}_A) < 0$, and therefore
$S_{i^\star}^\alpha = \emptyset$, for some $i^\star$ if
\begin{equation*}
  h_X(\bar{v}_A) < - \left[ (1-2\alpha) \frac{1 - \rho_A^{i^\star-1}}{1 - \rho_A} - \alpha \rho_A^{i^\star-1}\right] h_{BU}(\bar{v}_A),
\end{equation*}
which, given that $h_{BU}(\bar{v}_A) > 0$, may be recast as
\begin{equation*}
\alpha \rho_A^{i^\star-1} - (1-2\alpha) \frac{1 - \rho_A^{i^\star-1}}{1 - \rho_A} > H_{XU}(\bar{v}_A).
  \end{equation*}
  Rearranged for $\alpha$:
  \begin{equation*}
    \alpha > \frac{1 + H_{XY}(\bar{v}_A)(1 - \rho_A) - \rho_A^{i^\star - 1}}{2 - \rho_A^{i^\star-1} - \rho_A^{i^\star}} \ \text{when} \ \rho_A \neq 1.
    \end{equation*}
    Finally, we consider the case that $\rho_A = 1$. Then,
    \begin{equation*}
      h_{S_i^\alpha}(\bar{v}_A) \leq h_X(\bar{v}_A) + \left[ (1-2\alpha)(i-1)   - \alpha \right] h_{BU}(\bar{v}_A).
      \end{equation*}
      The set $S_{i^\star}^\alpha $ is empty for some $i^\star$ if
      \begin{equation*}
        \alpha > \frac{H_{XY}(\bar{v}_A) + i^\star -1}{2i^\star - 1} \ \text{when} \ \rho_A = 1. \qed
      \end{equation*}
          
\bibliography{extracted.bib}
          
\end{document}